\documentclass[aps,prl,twocolumn,groupedaddress, showpacs]{revtex4}

\usepackage{graphicx,amsmath,amssymb}
\usepackage{dcolumn}
\usepackage{bm}

\newcommand{\lp}{\ell_{\rm p}}
\newcommand{\lc}{\ell_{\rm c}}
\newcommand{\lpf}{\ell_{\rm p, floppy}}
\newcommand{\cs}{c_{\rm s}}
\newcommand{\cf}{c_{\rm f}}

\usepackage{color}
\renewcommand\mark[1]{\bgroup\color{red}{#1}\egroup}
\newcommand\remove[1]{\bgroup\color{blue}\bfseries{#1}\egroup}
\renewcommand\remove[1]{\bgroup\color{blue}{}\egroup}

\begin{document}


\title{Semiflexible Filamentous Composites}
\author{E.M. Huisman$^{1}$, C. Heussinger$^{2}$, C. Storm$^{3}$, G.T. Barkema$^{1,4}$\\[3ex]
  \normalsize{$^1$Universiteit Leiden, Instituut-Lorentz, Postbus 9506, NL-2300 RA Leiden, The Netherlands}\\
  \normalsize{$^2$Universit\'e de Lyon; Univ. Lyon I, Laboratoire de Physique de
    la Mati\`ere
    Condens\'ee et Nanostructures; UMR CNRS 5586, 69622 Villeurbanne, France}\\
  \normalsize{$^3$Department of Applied Physics and Institute for Complex
    Molecular Systems, Eindhoven University of
    Technology, P. O. Box 513, NL-5600 MB Eindhoven, The Netherlands}\\
  \normalsize{$^4$Universiteit Utrecht, Institute for Theoretical Physics,
    NL-3584 CE Utrecht, The Netherlands} }

\begin{abstract}
Inspired by the ubiquity of composite filamentous networks in nature we
investigate models of biopolymer networks that consist of interconnected
floppy and stiff filaments. Numerical simulations carried out in three
dimensions allow us to explore the microscopic partitioning of stresses and
strains between the stiff and floppy fractions $\cs$ and $\cf$, and reveal a
non-trivial relationship between the mechanical behavior and the relative fraction of
stiff polymer: when there are few stiff polymers, non-percolated stiff
``inclusions`` are protected from large deformations by an encompassing floppy
matrix, while at higher fractions of stiff material the stiff network is
independently percolated and dominates the mechanical response.
\end{abstract}

\maketitle

The basic design of most structural biological materials is that of a
crosslinked meshwork of semiflexible protein polymers. The mechanical properties
of these biomaterials are biologically highly
significant~\cite{Bausch,Weitz}.  Understanding these properties at the bulk or
continuous level is not sufficient: biological entities like cells, motor- and
sensing proteins experience, manipulate and interact with these polymer networks
at single-filament lengthscales and are therefore intimately aware of the
discrete nature of these materials. Another complication arises when considering
that most structural biomaterials are in fact composites: bi- or polydisperse
mixtures of different protein polymers. The extracellular matrix (ECM) consists
of a mixture of stiff collagen and flexible elastin filament (bundles), and the
relative abundance of these two greatly affects mechanical
properties~\cite{Black}. A more specific example that derives much of its biological function from the side-by-side
deployment of mechanically vastly different filaments is articular
cartilage - a complex, partially orded composite containing type-II
collagen and proteoglycans as its main structural components \cite{Bullough}. Composite physics may be at play in single-component
networks: coexistent and interlinked single fibers and fiber bundles determine
the mechanical properties of actin gels and actin-filamin
networks~\cite{Schmoller}. The interplay between stiff and floppy
elements goes far beyond simple property mixing: The network of relatively
floppy f-actin and intermediate filaments is believed to be nonlinearly
stiffened by the rigid microtubules, and experiments have hinted at significant
tensional forces in the cellular actin~\cite{Ingber,Mizushima}. The cell
cytoskeleton, that is built up from microtubules, actin filaments and
intermediate filaments, is yet another striking example of a composite network.

Significant effort has been devoted to model systems of homogeneous and
isotropic {\it single-component} networks of biopolymers, such as f-actin and
collagen~\cite{MacKintosh,Storm,Janmey,Janmey2,Heussinger,Timonen,Huisman,Onck, Huisman2}. The single filaments that constitute these networks can be described
by the semiflexible wormlike chain force-extension curve, where extension
requires that thermal fluctuations of the filaments be suppressed leading to a
steep and non-linear increase in the force. Compression requires considerably
smaller forces that become constant in the Euler buckling
limit~\cite{MacKintosh,Storm}.  Networks of such filaments show highly
non-linear strain-stiffening and negative normal forces under
shear~\cite{Janmey,Janmey2}.  Recent theoretical studies and 
simulations have also underlined the importance of non-affine bending
deformations in these
networks~\cite{Heussinger,Huisman,Onck, Huisman2}. Models applying a similar
method to composite biomaterials have only recently begun to
emerge~\cite{Broedersz, Tanaka} and have focused on bulk behavior.

In this Letter, we report the results of a series of numerical experiments of
{\it two-component} networks of biopolymers to determine the relationship
between composition and mechanical properties, both on the single-filament as
well as on the bulk level. Furthermore, we compare our results to a theoretical
model.
\begin{figure*}[t]
\begin{center}
\includegraphics[angle=0,width=1.0\linewidth]{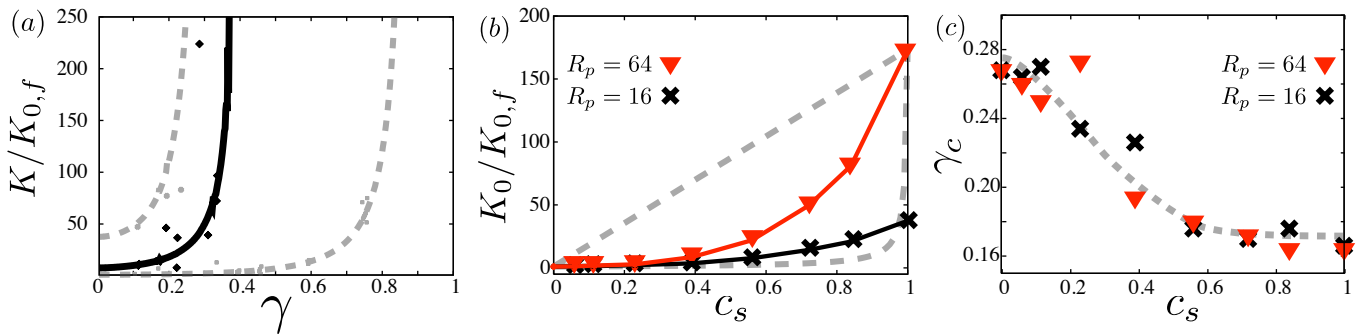}
\end{center}
\caption{Macroscopic properties of the networks as a function of the fraction
  $c_s$ of stiff filaments in a network. (a) Shear modulus $K$ as a function of
  shear $\gamma$, normalized by the initial shear modulus $K_{0,f}$ of
  single-component networks of floppy filaments. The different curves represent
  the stiffness of networks with $c_s=0.0,0.56,1.0$ (from
    bottom to top), at a fixed persistence length ratio $R_p=16$. Some data
  points ($<1\%$) lie well outside the curve; these are indicated by the
  symbols. These outliers occur due to local reorientations, cf.~\cite{Huisman}.
  (b) The normalized initial shear modulus as a function of $c_s$,
  for networks with $R_p=64$ and 16. For comparison, we also plot
  curves corresponding to a linear scaling of the shear modulus with $c_s$,
  given by $K_0(c_s)=c_s K_{0,s} + (1-c_s)K_{0,f}$, and a linear scaling of the
  compliance with $c_s$, given by $1/K_0(c_s)=c_s/K_{0,s} + (1-c_s)/K_{0,f}$.
  (c) Critical shear $\gamma_c$, defined as the shear at which the shear modulus
  is twice the initial shear modulus, as a function of $c_s$. The curve is drawn
  as a guide to the eye.}
\label{fig1}
\end{figure*}

Our networks consist of long filaments that are permanently crosslinked.
These crosslinks force a binary bond between two
filaments, without angular preferences. The filaments are described by the
semiflexible wormlike chain model~\cite{MacKintosh,Storm}. The segments (parts
of filaments in between two crosslinks) have degrees of freedom in addition to
the curvilinear length and the locations of both ends, that are integrated to
obtain an effective Hamiltonian. This effective Hamiltonian gives the energy for
each network configuration, characterized by the network topology, the positions
of the crosslinks, and the link-to-link separation of the filaments between
crosslinks.  Starting from a random, isotropic network consisting of crosslinks
and segments, we apply a large number of Monte Carlo moves which alter the
network topology such that filaments with a persistent directionality along
segments are formed.  At this point, we designate filaments to be either stiff
or floppy by assigning to each segment a persistence length and an equilibrium
backbone length. We then further relax the configuration by applying new Monte
Carlo moves. All our networks have periodic boundary conditions and contain 1,000
crosslinks. Their lateral sizes are determined by the condition of
zero pressure. A detailed description of this approach is presented in~\cite{Huisman}.
Our networks are characterized by the following set
of parameters: the persistence length $\lp$ of the stiff filaments, the
stiffness ratio $R_p=\lp/\lpf$, the average filament length $L$, the average
distance between crosslinks along a polymer's backbone $\lc$ and finally the
relative fraction of stiff filaments $\cs$.  In the present work, we examine the
$\cs$-dependence of the mechanical behavior. We restrict ourselves to a
biologically relevant region of parameter space: the persistence length $\lpf$
of the floppy filaments and the crosslink distance $\lc$ are of comparable
magnitude. On average, each filament is crosslinked six times ($L=6l_c$). The
ratio of the persistence lengths of stiff and floppy filaments $R_p$ is chosen
to be 16 or 64. While this ratio is smaller than that for collagen/elastin ($R_p
\approx 100$)~\cite{Black} or microtubules/f-actin ($R_p>200$)~\cite{Gittes}, it
is large enough to capture the qualitative behavior of such composite networks.
Unless otherwise stated, all data shown represent the averages of nine network
realizations.

Our key findings are summarized in Fig.~\ref{fig1}. In Fig.~\ref{fig1}a we plot
a 50/50 composite: rather than averaged, the mechanical behavior is bimodal --
approaching the fully floppy system at low strains but, at finite strains,
resembling the fully stiff network. We stress that this type of response can
only be achieved in a composite. Even the linear behavior does not interpolate
simply between stiff and floppy: Fig.~\ref{fig1}b shows that at low to
intermediate $\cs$, the modulus is quite insensitive to $\cs$, but rises very
quickly at higher $\cs$. Fig.~\ref{fig1}c, finally, reinforces the point of
Fig.~\ref{fig1}a: although the effects of adding stiff polymer are hardly
noticeable in the linear elastic behavior, their effect on the nonlinear
behavior is felt much earlier. The critical strain ($\gamma_c$) for the
onset of the nonlinear regime reacts immediately to the addition of stiff
material, but saturates at a point roughly coincident with the rise of the
linear modulus.

The qualitative picture that emerges at small $\cs$ is
one of a floppy matrix encompassing isolated stiff filaments, or non-percolated
clusters. Intuitively, the initial insensitivity of the linear modulus
$K_0$ to the addition of stiff material makes sense: deforming the stiff
filaments requires higher energies than deforming the softer elements, and
therefore the low-energy modes of the system favor straining the floppy elements
over the stiff ones.  As long as stiff filaments do not form an independently
load-bearing subnetwork, these low-energy
modes exist and are compatible with the bulk deformation.

\begin{figure*}[t]
\begin{center}
\includegraphics[angle=0,width=1.0\linewidth]{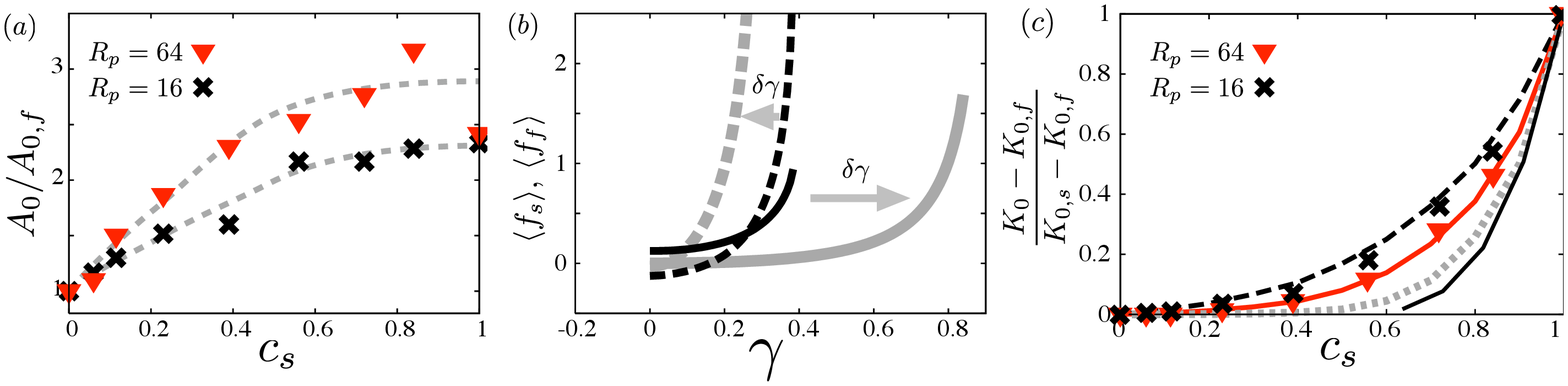}
\end{center}
\caption{{(a) The non-affinity at zero shear, divided by the initial
    non-affinity of a network with $c_s=0$. The curves
    are drawn as guides to the eye. (b) Average forces in the floppy and stiff
    filaments during deformation, shown by the solid and dotted curves, respectively.
    The black curves represent the average force of a network with
    $c_s=0.56$ during shear. For comparison, we plot the average forces in
    single-component networks, $c_s=1.0$ and $c_s=0.0$ (grey curves). As
    indicated by the arrows, the curves for the average forces in the composite
    networks are shifted along $\gamma$ with respect to the curves for the
    single-component network. (c) The scaled initial stiffness as a function of
    $c_s$, obtained by the floppy-mode model for
    $R_p=16,64,1000,\infty$ (from top to bottom). For comparison, the data from simulations are given
    by the symbols.}}
\label{fig2}
\end{figure*}

This interpretation is confirmed by an examination of the microscopic
deformation field, characterized by the non-affinity parameter $A =\langle
|\textbf{x}_{\mbox{aff}} - \textbf{x}|^2\rangle /\gamma^2$ (Fig.~\ref{fig2}a).
This parameter quantifies the deviation of the local deformations,
$\textbf{x}$, from a homogeneous affine deformation field,
$\textbf{x}_{\mbox{aff}}$. As shown in previous numerical work~\cite{Huisman2}
and in experiments~\cite{Wen}, the non-affinity generally increases with increased stiffness of the filaments,
as bending deformations are more important for the network response of stiffer filaments.
Indeed we find that the non-affinity is minimal
for purely floppy networks and rises roughly linearly with the addition of
stiff material.  Such a linear increase represents the generic behavior of
low-density (stiff) inclusions that independently perturb the deformation
field of their surrounding (floppy) matrix~\cite{Torquato}. These
additional non-affine deformations bring the floppy filaments closer to
the nonlinear part of their force-extension relation, giving rise to the
decrease of the critical strain $\gamma_c$ as displayed in Fig.~\ref{fig1}c.

In the nonlinear regime the inherent stiffening of a single
semiflexible polymer makes the distinction between floppy and stiff fractions
highly strain-dependent, with the ratio of their nonlinear moduli tending to unity
in the high strain limit. This suggests a self-matching behavior at finite
strain: the floppy network stiffens up to the point where its modulus matches that
of the stiff network. Beyond this point, the entire meshwork behaves as a nearly
monodisperse system of stiff filaments. This effect is the origin of the
behavior in Fig.~\ref{fig1}a: at high strains, the entire system is ultimately
forced to couple to the stiffer deformation modes.

This mechanism of stiffness matching is illustrated in Fig.~\ref{fig2}b which
shows the average forces in the stiff and floppy filaments during deformation.
By comparing with the one-component networks (grey lines) we can define a strain
shift $\delta\gamma$: For given network strain $\gamma$, the filaments in the
composite behave as if they were strained up to $\gamma+\delta\gamma$.
Apparently, the effective strain on the floppy filaments is much larger than that
on the stiff filaments. Equivalently, high forces in stiff
filaments are suppressed, at the cost of
increased forces in floppy filaments.

Interestingly, this load-partitioning persists even at \emph{zero
  strain}, where stiff filaments are, on average, compressed while floppy
filaments are stretched out. This stretched/compressed ground state is
tantalizingly reminiscent of tensegrity states~\cite{Ingber}. Apparently, dense crosslinking restricts relaxation of the network, and the absolute minimum
of mechanical energy cannot be attained. There is, therefore, always a finite
amount of residual elastic energy. This suggests that such force distributions
may not be a deliberate design principle but rather are the necessary byproduct
of polydispersity in filamentous composites.

The picture of a floppy matrix embedding stiffer inclusions breaks down when the
stiff filaments become independently rigidity percolated: the point where
deformation of the stiff elements becomes inevitable. We may estimate the percolation threshold by a counting
argument~\cite{Maxwell}. Equating the number of degrees of freedom of the stiff
filaments to the number of constraints due to crosslinks between stiff filaments
gives (for $L/\lc=6$) a threshold $\cs=0.56$.  This marks the transition from
the low to the high $\cs$ regime and coincides roughly with the rise in the linear
modulus $K_0$ (Fig.~\ref{fig1}b). Two separate obervations confirm the onset of stiff dominance:
Firstly, $\cs=0.56$ is the point at which the non-affinity, which we attribute
to the floppy matrix attempting to work around the stiff fraction, begins to
plateau at the level of the bending dominated response of a
purely stiff network. Secondly, the critical strain $\gamma_c$ levels off
around this same value of $\cs$.
In the range of stiffness ratios ($R_p$) accessible to the simulations the
percolation is rather ``soft'' and represents a smooth cross-over phenomenon. The approach towards the
singular percolation limit, $R_p=\infty$, has for example been studied in
simulations of mixed random resistor networks \cite{Straley}. To address the
analogous problem we compare our results with theoretical considerations,
in which the parameter $R_p$ can be tuned to arbitrarily large values. The
``floppy-mode'' theory~\cite{Heussinger} has recently been shown to capture
quite well the elasticity in one-component isotropic~\cite{Heussinger2,Lieleg}
as well as anisotropic networks~\cite{Missel}.  Within this theoretical
framework the calculation of the network elastic modulus is reduced to the
description of a ``test'' filament in an array of pinning sites.  The coupling
strength to these sites, $k$, represents the elastic modulus of the network and
has to be calculated self-consistently. To generalize this model to the case of
composite networks we use two different test chains with coupling parameters
$k_f$ and $k_s$, representing floppy and stiff filaments, respectively
\footnote{Details of the calculation will be published elsewhere.}. The use of
two different coupling strengths quite naturally takes into account the load
partitioning encountered in the simulations. The network modulus,
$k=c_sk_s+(1-c_s)k_f$, is obtained by solving the two equations
\begin{equation}\label{eq:selfconsEn}
  k_{f/s} \simeq  \left\langle \min_{y}
    \left( W_b^{f/s}\left[y(s)\right] + \frac{1}{2}\sum_{i=1}^n k_{\alpha_i}\left(y(s_i)-\bar
y_i\right)^2\right)\right\rangle     \,,
\end{equation}
where $k_{\alpha_i}=k_s,k_f$ with probability $c_s$ and $1-c_s$, respectively.
The two energy contributions on the rhs of Eq.~(\ref{eq:selfconsEn}) reflect the
competition between the bending energy of the (floppy or stiff) filament,
$W_b^{f/s}$, and the energy due to deformation of the surrounding medium by
displacing the pinning sites (located at arclength position $s_i$ along the
filament). The nonlinear entropic stretching elasticity is not included in
  these equations. The minimization is to be performed over the contour of the
filament, $y(s)$, the angular brackets specify the disorder average over the
network structure.

Fig.~\ref{fig2}c displays the results from this calculation for various
stiffness ratios $R_p$, showing a sharp percolation transition in the
limit $R_p\to\infty$.  The model compares well with the simulation
data, even though the stretching elasticity is not accounted
for. This indicates that the bending stiffness is likely the dominant
factor in determining the rise of the linear elastic modulus, in agreement
with the proposed mechanism of load-partitioning and the observed increase
of the non-affinity.

In conclusion, our results demonstrate that the mechanical behavior of filamentous composites
is considerably richer than the simple proportional mixing of properties. The
fact that the floppy and stiff networks are physically linked causes a
strongly nonlinear coupling between the strain fields which deeply affects
composite mechanics. This may explain the ubiquity of composites in structural
biological applications: slight variations in composition cause large changes in
mechanical behavior. This high susceptibility makes the composite architecture
an attractive motif for biological regulation.  Likewise, the ``best of both
worlds`` aspect may be exploited by Nature: composites combine the initial
softness of their most compliant components with the ultimate toughness of the
stiffest elements. This greatly enhances the stiffness range of nonlinearly
elastic materials. Moreover, composites do so in a manner that could never be
attained in monodisperse materials, since linear and nonlinear properties of
composites are determined by two physically different materials and therefore
may be independently varied. This possibility of independently tuning the linear
and nonlinear behavior also has considerable potential for the
design of biomimetic or bio-inspired synthetic materials and deserves further
exploration.

It is a pleasure to thank Paul Janmey for helpful discussions. CS and CH
acknowledge the hospitality of the Aspen Center for Physics where part of this
work was carried out. CH thanks the von Humboldt foundation for financial
support.

\end{document}